\documentclass{article} 
\RequirePackage{amsfonts}       
\usepackage{amssymb}
\usepackage{amsthm}

\newcommand{\Omit}[1]{}

\newtheorem{definition}{Definition}
\newtheorem{theorem}{Theorem}
\newtheorem{corollary}{Corollary}
\newtheorem{lemma}{Lemma}

\newenvironment{remark}{\medskip\noindent \textsc{Remark.}} {\medskip}

\newenvironment{pf}{\em \medskip\noindent \textsc{Proof.}}
{\hspace*{\fill}\nolinebreak[2]\hspace*{\fill}$\blacksquare$\medskip}

\newbox\itembox
\def\itemlistlabel#1{#1\hfill}
\def\itemlist#1{\setbox\itembox=\hbox{#1}%
                \list{}{\labelwidth\wd\itembox
                             \leftmargin\labelwidth
                             \advance\leftmargin by\itemindent
                             \advance\leftmargin by\labelsep
                             \let\makelabel\itemlistlabel}}

\renewcommand{\phi}{\varphi}
\newcommand{\tuple}[1]{\langle #1 \rangle}

\newcommand{\card}[1]{\mathit{Card}(#1)}           %
\newcommand{\ext}[1]{|#1|}

\newcommand{\eqv}{\leftrightarrow}      
\newcommand{\imp}{\rightarrow}          
\newcommand{\subfml}{\mathit{sf}}
\newcommand{\lngth}[1]{|\!|#1|\!|}

\newcommand{\cstit}[1]{[{#1}]}           %
\newcommand{\poscstit}[1]{\langle {#1} \rangle}    %
\newcommand{\dstit}[2]{[{#1}\ \mathit{dstit}\! :{#2}]}

\newcommand{\InclBox}[1]{$\Box \!\! \imp \!\! #1$}      

\newcommand{\atmset}{\ensuremath{\mathit{ATM}}}        

\newcommand{\agtset}{\ensuremath{\mathit{AGT}}}        

\newcommand{\STIT} {{\textsf{STIT}}}              %
\newcommand{\CSTIT}{{\textsf{CSTIT}}}            %
\newcommand{\DSTIT}{{\textsf{DSTIT}}}            %

\newcommand{\LCSTIT}{$\mathcal{L}_{\mathsf{CSTIT}}^\agtset$}
\newcommand{\LDSTIT}{$\mathcal{L}_{\mathsf{DSTIT}}^\agtset$}

\title{Alternative axiomatics and complexity of deliberative \STIT\ theories}

\author{Philippe Balbiani, Andreas Herzig and Nicolas Troquard
\\ Institut de recherche en informatique de Toulouse (IRIT), France
\\ \small{\texttt{$\{$balbiani,herzig,troquard$\}$@irit.fr}}
}

\date{}

\begin{document}

\maketitle

\begin{abstract}
We propose two alternatives to Xu's axiomatization of the Chellas \STIT.
The first one also provides an alternative axiomatization of
the deliberative \STIT.
The second one starts from the idea that the historic necessity operator
can be defined as an abbreviation of operators of agency, and
can thus be eliminated from the logic of the Chellas \STIT.
The second axiomatization also allows us to establish
that the problem of deciding the satisfiability of a \STIT\ formula
without temporal operators is NP-complete in the single-agent case, and
is NEXPTIME-complete in the multiagent case, both for the deliberative
and the Chellas' \STIT.
\end{abstract}

\tableofcontents

\section{Introduction}

\STIT\ theory is one of the most prominent accounts of agency in
philosophy of action. It is the logic of constructions of the form
`agent $i$ sees to it that $\phi$ holds'.
While \STIT\ has played an important role in philosophical logic
since the 80ies, it seems to be fair to say that its mathematical aspects
have not been developed to the same extent. Most probably the reason is that
\STIT's models of agency are much more complex than
those existing for other modal concepts (such as say necessity, belief,
or knowledge): first, the `seeing-to-it-that' modalities interact
(or perhaps better: must be guaranteed not to interact)
because the agents' choices are supposed to be independent;
second there is another kind of modality involved, viz.\
the `master modality' of historic necessity.
There are also temporal modalities, but just as most of the other
proof-theoretic approaches to \STIT, we do not investigate
these here.

As a consequence, proof systems for \STIT\ are rather complex, too.
To our knowledge the following have been proposed in the literature.

\begin{itemize}
\item Xu provides Hilbert-style axiomatizations in terms of
the historic necessity operator and Chellas' \STIT\ operator
\cite[Chap.\ 17]{belnap01facing},
without considering temporal operators.
As the deliberative \STIT-operator can be expressed in terms of Chellas'
(together with the historic necessity operator),
the axiomatization transfers to the deliberative \STIT.
Xu proves their completeness (without considering the temporal dimension),
by means of canonical models, and proves decidability by means of filtration.
Besides, Xu also gives a complete axiomatization of the one-agent
achievement \STIT\ \cite[Chap.\ 16]{belnap01facing}.

\item Wansing provides a tableau proof system for the deliberative \STIT\ \cite{wansing06aiml}.
The system is complete, but does not guarantee
termination, and thus ``is not tailored for defining tableau
algorithms'' \cite{wansing06aiml}.

\item D\'{e}gremont gives a dialogical proof procedure for the
deliberative \STIT\ \cite{Degremont}. Again, the system is complete,
but does not guarantee termination, and can therefore
only be used to build proofs by hand.
\end{itemize}

In this note, we focus on the so-called Chellas \STIT\ named after his
proponent \cite{chellas69phd, chellas92agency}. The original operator
defined by Chellas is nevertheless notably different since it does not
come with the principle of independence of agents that plays a central
role here. Following its presentation in \cite{horty95deliberative},
we use the term \CSTIT\ to refer to the logic of that modal operator.
We show that Xu's axiomatics of the logic of the Chellas \STIT\ can be
greatly simplified. After recalling it (Section \ref{sec:background})
we propose an alternative one and prove its completeness  (Section \ref{sec:alterAx}).
Based on the latter we show that in presence of at least two agents,
the modal operator of historic necessity
can be defined as an abbreviation (Section \ref{sec:unsettling}).
This leads to a simplified semantics (Section \ref{sec:newSemantics}),
and to characterizations of the complexity of satisfiability
(Section \ref{sec:complexity}).

\section{Xu's axioms for the \CSTIT}
\label{sec:background}

Some preliminary remarks are due. In \cite[Chap.\ 17]{belnap01facing},
Ming Xu presents $Ldm$, an axiomatization for the basic
(that is, without temporal operators) \emph{deliberative} \STIT\ logic.
As pointed out, deliberative \STIT\ logic and Chellas' \STIT\ logic are
interdefinable and just differ in the choice of primitive operators.
Following Xu we refer to these two logics as the
\emph{deliberative \STIT\ theories}.
We here mainly focus on $Ldm$ with the Chellas \STIT\ operator as
primitive.

\subsection{Language}
The language of Chellas' \STIT\ logic is built from
a countably infinite set of atomic propositions $\atmset$ and
a countable set of agents $\agtset$.
To simplify notation we suppose that $\agtset $ is an initial subset
$\{0,1,\ldots\}$ of $\mathbb{N}$ (possibly $\mathbb{N}$ itself).

Formulas are built by means of the boolean connectives together
with modal operators of historic necessity and of agency in the standard way.
Usually these modal constructions are noted
$\mathit{Sett:}\ \phi $ (`$\phi$ is settled') and
$[i\ \mathit{cstit}\! :\phi]$ (`$i$ sees to it that $\phi$'),
where
$i \in \agtset$. For reasons of conciseness we here prefer to use
$\Box \phi$ instead of $\mathit{Sett:}\ \phi $,
and $\cstit{i} \phi $
instead of $[i\ \mathit{cstit}\! :\phi]$.
The language \LCSTIT\ of the Chellas \STIT\ is therefore defined by the following BNF:
$$\phi \ ::=\ p \mid \lnot \phi \mid (\phi \land \phi)
              \mid \cstit{i}\phi \mid \Box\phi$$
where $p$ ranges over $\atmset$ and $i$ ranges over $\agtset$.
This provides a standard notation for the dual constructions
$\Diamond \phi$ and $\poscstit{i} \phi $, respectively abbreviating
$\lnot \Box \lnot \phi$ and $\lnot \cstit{i} \lnot \phi $.

The language \LDSTIT\ of the deliberative \STIT\ is defined by:
$$\phi \ ::=\ p \mid \lnot \phi \mid (\phi \land \phi)
              \mid \dstit{i}\phi \mid \Box\phi$$
Note that neither \LCSTIT\ nor \LDSTIT\ contain temporal operators.

The following function will be useful to compute
the number of symbols that are necessary to write down $\phi$.

\begin{definition}
We define recursively a mapping $\lngth{.}$ from formulas of \LCSTIT $\cup$ \LDSTIT\
to $\mathbb{N}$ :
$\lngth{p}                 = 1              $,
$\lngth{\lnot\phi}         = 1+\lngth{\phi} $,
$\lngth{(\phi \land \psi)} = 3 + \lngth{\phi} + \lngth{\psi}$,
$\lngth{\cstit{i}\phi}   = 3 + \lngth{\phi}$, and
$\lngth{\dstit{i}{\phi}} = 5 + \lngth{\phi}$.
\end{definition}

\goodbreak
\subsection{Semantics}
The semantics of the \CSTIT\ is extensively studied in Belnap et al.\ \cite{belnap01facing}.
It consists of a branching-time structure (BT)
augmented by the set of agents and a choice function (AC).
Here, we refer to BT + AC models as \STIT-models.

A \emph{BT structure} is of the form $\tuple{W, < }$, where
$W$ is a nonempty set of moments, and
$<$ is a tree-like ordering of these moments: for any
$w_1$, $w_2$ and $w_3$ in $W$, if $w_1 < w_3$ and $w_2 < w_3$,
then either $w_1 = w_2$ or $w_1 < w_2$ or $w_2 < w_1$.

A maximal set of linearly ordered moments from $W$ is a \emph{history}.
When $w \in h$ we say that moment $w$ is \emph{on} the history $h$.
$Hist$ is the set of all histories.
$H_w = \{h | h \in Hist, w \in h\}$ denotes the set of histories passing through $w$.
An \emph{index} is a pair $w/h$, consisting of a moment $w$ and a
history $h$ from $H_w$ (i.e., a history and a moment in that history).

A \emph{BT+AC model} is a tuple $\mathcal{M} = \tuple{W,<,Choice,V}$, where:
\begin{itemize}
\item $\langle W,<\rangle$ is a BT structure;

\item $Choice : \agtset \times W \rightarrow 2^{2^{Hist}}$
is a function mapping each agent and each moment $w$ into a partition of $H_w$,
such that
    \begin{itemize}
    \item $Choice_i^w \not = \emptyset$;
    \item $Q \not = \emptyset$ for every $Q \in Choice_i^w$;
    \item for all $w $ and all mappings $s_w : \agtset \longrightarrow 2^{H_w}$
          such that $s_w(i) \in Choice_i^w$,
          we have $\bigcap_{i\in \agtset} s_w(i) \not = \emptyset$.
    \end{itemize}

\item $V$ is valuation function $V : \atmset \rightarrow 2^{W \times Hist}$.
\end{itemize}
The equivalence classes belonging to $Choice_i^w$ can be thought of as possible choices
that are available to agent $i$ at $w$.
Given a history $h \in H_w$, $Choice_i^w(h)$ represents the particular choice
from $Choice_i^w$ containing $h$, or in other words, the particular action
performed by $i$ at the index $w/h$.
We call the constraint of nonempty intersection of all possible simultaneous choices of agents
(or: strategy profile) the \emph{superadditivity constraint}.

A formula is evaluated with respect to a model and an index.\\

\begin{tabular}{lcl}
$\mathcal{M},w/h \models p$ & iff  & $w/h \in V(p), p \in \atmset$\\
$\mathcal{M},w/h \models \lnot \phi$ & iff & $\mathcal{M},w/h \not\models \phi$\\
$\mathcal{M},w/h \models \phi \land \psi$ & iff & $\mathcal{M},w/h \models \phi$ and
                                                  $\mathcal{M},w/h \models \psi$\\
$\mathcal{M}, w/h \models \Box \phi     $ & iff & $\mathcal{M}, w/h' \models \phi,
                                                                   \forall h' \in H_w$\\
$\mathcal{M}, w/h \models \cstit{i} \phi $ & iff & $\mathcal{M}, w/h'\models \phi,
                                                        \forall h' \in Choice_i^w(h)$\\

$\mathcal{M}, w/h \models \dstit{i}{\phi}$ & iff & $\mathcal{M}, w/h' \models \phi,
                                                         \forall h' \in Choice_i^w(h)$\\
                      && and $\exists h'' \in H_w,\ \mathcal{M}, w/h'' \models \lnot\phi$
\end{tabular}\\

Hence historical necessity (or inevitability) at a moment $w$ in a history is
truth in all histories passing through $w$.
According to Chellas, an agent $i$ sees to it that $\phi$ in a moment-history pair $w/h$
if $\phi $ holds on all histories that agree with $i$'s current choice.

\emph{Validity in BT+AC structures}
is defined as truth at every moment-history pairs of every BT+AC-models.
A formula $\phi$ is satisfiable in BT+AC structures if $\lnot \phi$ is not valid
in BT+AC structures.

The following valid equivalences justify the interdefinability of
our \STIT-operators:
\begin{center}
\begin{tabular}{lcl}
$\dstit{i}{\phi}$ & $\eqv$ & $\cstit{i} \phi \land \lnot\Box\phi $ \\
$\cstit{i} \phi $ & $\eqv$ & $\dstit{i}{\phi} \lor \Box\phi      $
\end{tabular}
\end{center}

\goodbreak
\subsection{Axiomatics}
Xu gave the following axiomatics of Chellas' \CSTIT:
\begin{itemlist}{(AAIA$_k$)}
  \item[S5($\Box$)]     the axiom schemas of S5  for $\Box$
  \item[S5($i$)]        the axiom schemas of S5 for every $\cstit{i}$
  \item[(\InclBox{i})] $\Box \phi \imp \cstit{i} \phi  $
  \item[(AIA$_k$)]
  $(\Diamond \cstit{0} \phi_0 \land \ldots \land
    \Diamond \cstit{k} \phi_k                    ) \imp
    \Diamond (\cstit{0} \phi_0 \land \ldots \land
              \cstit{k} \phi_k                   )  $
\end{itemlist}
The last item is a family of \emph{axiom schemes for
independence of agents} that is parameterized by the integer $k$.%
\footnote{
Xu's original formulation of (AIA$_k$) is
$$
(\mathit{diff}(i_0,\ldots,i_k) \land
    \Diamond \cstit{i_0} \phi_0 \land \ldots \land
    \Diamond \cstit{i_k} \phi_k                    ) \imp
    \Diamond (\cstit{i_0} \phi_0 \land \ldots \land
              \cstit{i_k} \phi_k                   )
$$       
for $1 \leq k$.
The difference predicates $\mathit{diff}(i_0,\ldots, i_k)$ express that
$i_0,\ldots, i_k $ are all distinct.
They are defined from an equality predicate $=$ whose domain is $\agtset$.
Formally we have to add the axioms:
$\mathit{diff}(i_0) \eqv \top$, and
\\ \centerline{
$\mathit{diff}(i_0,\ldots,i_{k+1}) \eqv
 \mathit{diff}(i_0,\ldots,i_k)
 \land i_1 \not = i_{k+1} \land \ldots  \land i_k \not = i_{k+1} $.
}
In consequence Xu's axiomatics has to contain axioms for equality.
We here preferred not to introduce equality in order to stay with the same
logical language throughout.

Clearly, each of our (AIA$_k$) can be proved from Xu's original (AIA$_k$).
The other way round, given $k$ and pairwise different $i_0,\ldots,i_k$,
suppose w.l.o.g.\ that $i_k \geq i_n$ for $n \leq k$.
Then one can prove Xu's (AIA$_k$)
\\ \centerline{
  $(\Diamond \cstit{i_0} \phi_{i_0} \land \ldots \land
    \Diamond \cstit{i_k} \phi_{i_k}                    ) \imp
    \Diamond (\cstit{i_0} \phi_{i_0} \land \ldots \land
              \cstit{i_k} \phi_{i_k}                   )   $
}
from our (AIA$_{i_k}$)
\\ \centerline{
  $(\Diamond \cstit{0} \phi_0 \land \ldots \land
    \Diamond \cstit{i_k} \phi_{i_k}                    ) \imp
    \Diamond (\cstit{0} \phi_0 \land \ldots \land
              \cstit{i_k} \phi_{i_k}                   )  $
}
by appropriately choosing $\phi_{n}$ to be $\top$ for all those $n < i_k$
that are not among $i_0,\ldots,i_k$: as $\cstit{n} \phi_n \eqv \top$
and $\Diamond \cstit{n} \phi_n \eqv \top$ hold, these conjuncts can be dropped
from our (AIA$_{i_k}$).

}

\begin{remark}
As (AIA$_{k+1}$) implies (AIA$_k$), the family of schemas can be replaced by
the single (AIA$_{\card{\agtset}-1}$) when $\agtset$ is finite.
\end{remark}

Xu's system has the standard inference rules of modus ponens and
necessitation for $\Box$. From the latter necessitation rules for
every $\cstit{i}$ follow by axiom (\InclBox{i}).

\begin{theorem}[\mbox{\cite[Chapter 17]{belnap01facing}}]\label{XuCompleteness}
A formula $\phi$ of \LCSTIT\ is valid in BT+AC structures iff $\phi$ is provable from
the schemas S5($\Box$), S5($i$), (\InclBox{i}), and (AIA$_k$) by the rules of
modus ponens and $\Box$-necessitation.
\end{theorem}

Xu's decidability proof proceeds by building a canonical model
followed by filtration \cite[Theorems 17-18]{belnap01facing}.
Although he does not mention complexity issues,
when decidability is proved by canonical model construction
from which a finite model is obtained by filtration, then
``a NEXPTIME algorithm is usually being employed''
\cite[Appendix C, p.\ 515]{Blackburn:2001:ML}.
Therefore it can be expected that the problem of deciding
the satisfiability of a given formula of \LCSTIT\
is in NEXPTIME.
We shall characterize complexity precisely in Section \ref{sec:complexity}.

\goodbreak
\section{An alternative axiomatics }\label{sec:alterAx}

We now prove that (AIA$_k$) can be replaced by the family of axiom schemes
\begin{itemlist}{(\InclBox{i})}
  \item[(AAIA$_k$)]
  $ \Diamond \phi \imp
    \poscstit{k} \bigwedge_{0 \leq i < k} \poscstit{i} \phi  $   \hfill for $k \geq 1$
\end{itemlist}
We call (AAIA$_k$) the \emph{alternative axiom schema for independence of agents}.
Just as Xu's (AIA$_k$), (AAIA$_k$) involves $k+1$ agents.

\begin{lemma}[validity of AAIA$_k$]\label{lem:validAaia}
For each $k \geq 1$,
$\Diamond\phi \imp \poscstit{k} \bigwedge_{0 \leq i < k} \poscstit{i}\phi$
is valid in BT+AC structures.

\begin{pf}
See Annex.
\end{pf}
\end{lemma}

To warm up, we first prove that our (AAIA$_1$) implies Xu's (AIA$_1$).

\begin{lemma}\label{ortho-theo}
The schema (AIA$_1$) is provable from
S5($\Box$), S5($i$), (\InclBox{i})
and:
\begin{itemlist}{(\InclBox{i})}
  \item[(AAIA$_1$)]
$\Diamond\phi \imp \poscstit{1} \poscstit{0} \phi $
\end{itemlist}
by modus ponens and $\Box$-necessitation.

\begin{pf}
We establish the following deduction:
\begin{enumerate}
 \item $
\Diamond \cstit{0} \phi_0 \imp \poscstit{1} \poscstit{0}\cstit{0} \phi_0
$ \hfill from axiom (AAIA$_1$), substituting $\cstit{0}\phi_0$ for $\phi$
\item $
\Diamond \cstit{0} \phi_0 \imp \poscstit{1} \cstit{0} \phi_0
$ \hfill from previous line by S5($0$)
\item $
\Diamond \cstit{0} \phi_0 \land \cstit{1}\phi_1 \imp \poscstit{1} \cstit{0} \phi_0 \land \cstit{1}\cstit{1}\phi_1
$    $~$\hfill from previous line by S5($1$)
\item $
\Diamond \cstit{0} \phi_0 \land \cstit{1}\phi_1 \imp \poscstit{1}(\cstit{0} \phi_0 \land \cstit{1}\phi_1 )
$ 
\hfill from previous line by K($1$)
\item $
\Diamond (\Diamond \cstit{0}\phi_0 \land \cstit{1} \phi_1) \imp \Diamond\poscstit{1} (\cstit{0}\phi_0 \land \cstit{1} \phi_1)
$ \\
$~$ \hfill from previous line by $\Box$-necessitation and K($\Box$)
\item $
\Diamond\cstit{0}\phi_0 \land \Diamond \cstit{1} \phi_1 \imp \Diamond\poscstit{1} (\cstit{0}\phi_0 \land \cstit{1} \phi_1)
$ 
\hfill from previous line by S5($\Box$)
\item $
\Diamond\cstit{0}\phi_0 \land \Diamond \cstit{1} \phi_1 \imp \Diamond (\cstit{0}\phi_0 \land \cstit{1} \phi_1)
$ \\    $~$
\hfill from previous line by (\InclBox{i}) axiom
and S5($\Box$)

\end{enumerate}
\end{pf}

\end{lemma}

We turn back to an arbitrary number of agents.

\begin{lemma}
\label{provable-lemma}
Every schema (AIA$_k$) is provable from
S5($\Box$), S5($i$), (\InclBox{i})
and (AAIA$_k$)
by the rules of modus ponens and $\Box$-necessitation.

\begin{pf}
We proceed by induction on $k$. The base case $k = 1$ is settled
by \mbox{Lemma \ref{ortho-theo}}. Now, suppose AIA$_{k-1}$ is
provable:
$$\Diamond\cstit{0}\phi_0 \land
                   \ldots \land \Diamond\cstit{k-1}\phi_{k-1} \imp
   \Diamond(\cstit{0}\phi_0 \land \ldots \land \cstit{k-1}\phi_{k-1}).$$
We prove AIA$_k$ with the following steps.
\begin{enumerate}
\item 
$\bigwedge_{i < k}\Diamond\cstit{i}\phi_i \imp
\Diamond\bigwedge_{i < k}\cstit{i}\phi_i$ $~$ \hfill by induction
hypothesis (AIA$_{k-1}$)
\item 
$\bigwedge_{i < k}\Diamond\cstit{i}\phi_i \imp
\poscstit{k}(\bigwedge_{j <
k}\poscstit{j}\bigwedge_{i<k}\cstit{i}\phi_i)$ $~$ \hfill from
previous line by (AAIA$_{k}$)
\item 
$\bigwedge_{i < k}\Diamond\cstit{i}\phi_i \imp
\poscstit{k}\bigwedge_{j < k}\poscstit{j}\cstit{j}\phi_j$ $~$
\hfill from previous line by K($j$)
\item 
$\bigwedge_{i < k}\Diamond\cstit{i}\phi_i \land
                   \cstit{k}\phi_k     \imp
\poscstit{k}(\bigwedge_{j < k}\cstit{j}\phi_j)
                   \land \cstit{k}\phi_k$\\
$~$ \hfill from previous line by S5($i$)
\item 
$\bigwedge_{i < k}\Diamond\cstit{i}\phi_i \land
                   \cstit{k}\phi_k     \imp
\poscstit{k}\bigwedge_{j \leq k}\cstit{j}\phi_j $ $~$ \hfill from
previous line by S5($k$)
\item 
$\Diamond(\bigwedge_{i < k}\Diamond\cstit{i}\phi_i \land
                   \cstit{k}\phi_k)     \imp
\Diamond\poscstit{k}\bigwedge_{j \leq k}\cstit{j}\phi_j $\\
$~$ \hfill from previous line by $\Box$-necessitation and K($\Box$)
\item 
$\Diamond(\bigwedge_{i < k}\Diamond\cstit{i}\phi_i \land
                   \cstit{k}\phi_k)     \imp
\Diamond\bigwedge_{j \leq k}\cstit{j}\phi_j $\\
$~$ \hfill from previous line by (\InclBox{i}) axiom and S5($\Box$)
\item 
$\bigwedge_{i \leq k}\Diamond\cstit{i}\phi_i \imp
\Diamond\bigwedge_{j \leq k}\cstit{j}\phi_j $ $~$ \hfill from
previous line by S5($\Box$)
\end{enumerate}

\end{pf}

\end{lemma}

\begin{theorem}
A formula of \LCSTIT\ is valid in BT+AC structures iff it is provable from the
axiom schemas S5($\Box$), S5($i$), (\InclBox{i}) and (AAIA$_k$) by the rules
modus ponens and $\Box$-necessitation.

\begin{pf}
First, observe that Xu's axiomatics and ours only differ by the schemas
(AIA$_k$) and (AAIA$_k$).

Soundness follows from:
\begin{enumerate}
\item the validity of our schemas AAIA$_k$ (see Lemma \ref{lem:validAaia}),
\item the validity of the rest of the axioms, and
\item the fact that modus ponens and $\Box$-necessitation preserve validity.
\end{enumerate}
The last two points are warranted by the soundness of Xu's axioms
(Theorem~\ref{XuCompleteness}).

Completeness follows from provability of Xu's (AIA$_k$) from our (AAIA$_k$)
(see Lemma \ref{provable-lemma}).
As observed above, the rest of Xu's axioms is directly present in our axiomatics.
\end{pf}

\end{theorem}

An alternative axiomatics of the deliberative \STIT\ is obtained viewing
$\cstit{i} \phi$ as an abbreviation of $\dstit{i} \phi \lor \Box \phi$.

\goodbreak
\section{Historic necessity is superfluous
         in presence of two agents or more}\label{sec:unsettling}

In this section, we suppose that $\card{\agtset} \geq 2$, i.e.\
there are at least agents $0$ and~$1$.

The equivalence $\Diamond \phi \eqv \poscstit{1} \poscstit{0} \phi  $
is provable from (AAIA$_1$), (\InclBox{i}) and S5($\Box$).
This suggests that $\Box \phi$ can be viewed as an abbreviation of
$\cstit{1} \cstit{0} \phi $.
Let us take this as an axiom schema.

\begin{itemlist}{(\InclBox{i})}
  \item[Def($\Box$)]
  $\Box \phi \eqv \cstit{1} \cstit{0} \phi $
\end{itemlist}

Pushing this further we can prove that under Def($\Box$), axiom (AAIA$_k$) can be
replaced by the family of axiom schemas of general permutation:
\begin{itemlist}{(\InclBox{i})}
\item[(GPerm$_k$)]
$\poscstit{l}\poscstit{m}\phi \imp
\poscstit{n}\bigwedge_{i\leq k, i \not = n} \poscstit{i}\phi$    \hfill for $k \geq 0$
\end{itemlist}
Note that similar to Xu's axiomatization, if $\agtset$ is finite then the single
schema (GPerm$_{\card{\agtset}-1}$) is sufficient.

The next lemma establishes soundness.

\begin{lemma}\label{lem:validGperm}
(GPerm$_k$) is valid in BT+AC structures.

\begin{pf}
See Annex.
\end{pf}
\end{lemma}

Now we prove that the principles of the preceding section can be derived.

\begin{lemma}\label{lem:BoxIsS5}
The axiom schemas of S5($\Box$),
and the schemas (\InclBox{i}) and (AAIA$_k$)
are provable from Def($\Box$), S5($i$) and (GPerm$_k$)
by the rules of modus ponens and $\cstit{i}$-necessitation,
and $\Box$-necessitation is derivable.

\begin{pf}
First let us prove that the logic of $\Box$ is S5.
Clearly the K-axiom
$\Box(\phi\imp\psi) \imp (\Box\phi\imp\Box\psi) $
is provable using standard modal principles, and the T-axiom
$\Box\phi\imp\phi$ follows from S5($0$) and S5($1$).
It remains to prove the 5-axiom $\Diamond\phi \imp \Box\Diamond\phi$:
\begin{enumerate}
\item
$                  \poscstit{1}\poscstit{0}\phi \imp
 \cstit{1}         \poscstit{1}\poscstit{0}\phi       $  \hfill by S5($1$);
\item
$\cstit{1}         \poscstit{1}\poscstit{0}\phi \imp
 \cstit{1}         \poscstit{0}\poscstit{1}\phi       $  \hfill by (GPerm$_1$)
                                                      and K($1$);
\item
$\cstit{1}         \poscstit{0}\poscstit{1}\phi \imp
 \cstit{1}\cstit{0}\poscstit{0}\poscstit{1}\phi       $  \hfill by S5($0$)
                                                      and K($1$);
\item
$\cstit{1}\cstit{0}\poscstit{0}\poscstit{1}\phi \imp
 \cstit{1}\cstit{0}\poscstit{1}\poscstit{0}\phi       $  \hfill by (GPerm$_1$);
\item
$                  \poscstit{1}\poscstit{0}\phi \imp
 \cstit{1}\cstit{0}\poscstit{1}\poscstit{0}\phi       $  \hfill from lines 1-4.
\end{enumerate}
Finally, $\Box$-necessitation is derivable
by applying first $0$-necessitation and then $1$-necessitation.

Concerning (AAIA$_k$) it is easy to see that under Def($\Box$)
it is an instance of (GPerm$_k$), for all $k \geq 1$.
It remains to prove (\InclBox{i}).
Let us show that $\poscstit{i}\phi \imp \poscstit{1}\poscstit{0}\phi$:
\begin{enumerate}
\item $\poscstit{i}\phi \imp \poscstit{i}\poscstit{j}\phi$             \hfill by S5($i$);
\item $\poscstit{i}\poscstit{j}\phi \imp \poscstit{1}\poscstit{0}\phi$ \hfill by (GPerm$_1$);
\item $\poscstit{i}\phi \imp \poscstit{1}\poscstit{0}\phi$ \hfill from lines 1-2.
\end{enumerate}

\end{pf}

\end{lemma}

\begin{theorem}\label{cool-theo}
Suppose $\card{\agtset} \geq 2$.
Then a formula of \LCSTIT\ is valid in BT+AC structures
iff it is provable from
S5($i$),        
Def($\Box$),    
and (GPerm$_k$)                 
by the rules of modus ponens and $\cstit{i}$-necessitation.
\end{theorem}

\begin{remark}
If $\agtset = \{0,1\}$ then the validities of \LCSTIT\ are axiomatized by Def($\Box$),
S5($1$), S5($2$), and
$\poscstit{1} \poscstit{0} \phi  \eqv \poscstit{0} \poscstit{1} \phi $.
Moreover, the Church-Rosser axiom
$\poscstit{0} \cstit{1} \phi \imp \cstit{1} \poscstit{0} \phi$.
can be proved from S5($1$), S5($2$) and (GPerm$_1$).
Therefore $\STIT$ logic with two agents is a so-called product logic,
alias a two-dimensional modal logic \cite{Marx99,GabbayEtAl03}.
Such product logics are characterized by the permutation axiom
$\poscstit{0} \poscstit{1} \phi \eqv \poscstit{1} \poscstit{0} \phi  $
together with the Church-Rosser axiom.
Hence the logic of the two-agent \STIT\ is nothing but the product
S5$^2$ = S5$\otimes$S5.
\end{remark}

\goodbreak
\section{A simpler semantics}\label{sec:newSemantics}

All axiom schemes are in the Sahlqvist class \cite{Blackburn:2001:ML},
and therefore have a standard possible worlds semantics.

\emph{Kripke models} are of the form
$M = \tuple{W, R, V} $, where
$W$ is a nonempty set of possible worlds,
$R$ is a mapping associating to every $i \in \agtset$ an equivalence relation $R_i$ on $W$,
and $V$ is a mapping from $\atmset$ to the set of subsets of $W$.
We impose that $R$ satisfies the following property:

\begin{definition}[general permutation property]
We say that $R$ satisfies the \emph{general permutation property} iff
for all $w,v \in W$ and for all $l,m,n \in \agtset$,
if $\tuple{w,v} \in R_l \circ R_m $ then there is $u \in W$ such that:
$\tuple{w,u} \in R_n $ and $\tuple{u,v}\in R_i $
for every $i \in \agtset \setminus \{n\}$.
\end{definition}

We have the usual truth condition:
$$M,w\models \cstit{i}\phi \mbox{ \ iff\ }
  M,u\models\phi  \mbox{ for every } u \mbox{ such that } \tuple{w,u} \in R_i$$
and the usual definitions of validity and satisfiability.

\begin{lemma}\label{lem:prop-rel}
For every $M = \tuple{W, R, V} $, and every $i,j \in \agtset$, $R$ satisfies
the following properties:
\begin{enumerate}
\item If $i \not = j$ then $R_i \circ R_j = R_1 \circ R_0 $.

\item $R_i \circ R_j $ is an equivalence relation for every $i,j \in \agtset$.

\item $ (\bigcup_{i \in \agtset} R_i)^*  =  R_0 \circ R_1 = R_1 \circ R_0 $.
\end{enumerate}

\begin{pf}
(1) follows from the validity of
$\poscstit{i}\poscstit{j}\phi \imp \poscstit{1}\poscstit{0}\phi$ (due to (GPerm$_0$)), and
the validity of
$\poscstit{1}\poscstit{0}\phi \imp \poscstit{i}\poscstit{j}\phi$ (due to (GPerm$_j$),
given that $i \not = j$).

(2) follows from (1) and the fact that the S5-axioms are valid for $\Box$
(see Lemma \ref{lem:BoxIsS5}).

In (3), the right-to-left inclusion
$R_0 \circ R_1 \subseteq (\bigcup_{i \in \agtset} R_i)^* $
follows from the inclusion
$R_0 \circ R_1 \subseteq (R_0 \cup R_1)^* $.
For the left-to-right inclusion
suppose $\tuple{w,v} \in (\bigcup_{i \in \agtset} R_i)^* $.
Hence there are $i_0,\ldots,i_k$ such that
$\tuple{w,v} \in R_{i_0} \circ \ldots \circ R_{i_k} $.
As all the $R_{i_l}$ are equivalence relations we may suppose w.l.o.g.\ that
$i_l \not = i_{l+1} $.
\begin{itemize}
\item If $k$ is odd then
$R_{i_0} \circ \ldots \circ R_{i_k} = (R_0 \circ R_1)^{k/2} $ by (1).
The latter is equal to $R_0 \circ R_1$ by (2).

\item If $k$ is even then
$R_{i_0} \circ \ldots \circ R_{i_k} = (R_0 \circ R_1)^{(k-1)/2} \circ R_{i_k}
                                    = (R_0 \circ R_1)           \circ R_{i_k} $ by (1) and (2).
The latter is equal to                $R_0 \circ R_1           \circ R_{0} $ again by (1),
and to $R_0 \circ R_0           \circ R_{1} $ by (2),
which is equal to $R_0 \circ R_{1} $ because $R_0 $ is an equivalence relation.
\end{itemize}
It follows that $ (\bigcup_{i \in \agtset} R_i)^*  \subseteq  R_0 \circ R_1 $.

\end{pf}
\end{lemma}

\begin{theorem}\label{theo:Sahlqvist}
A formula of \LCSTIT\ is valid in Kripke models satisfying the general permutation property
iff it is provable from
\begin{itemlist}{(GPerm$_k$)}
\item[S5($i$)]     the axiom schemas of S5 for every $\cstit{i}$
\item[Def($\Box$)] $\Box \phi \eqv \cstit{1} \cstit{0} \phi $
\item[(GPerm$_k$)] $\poscstit{l}\poscstit{m}\phi \imp
                    \poscstit{n}\bigwedge_{i\leq k, i \not = l} \poscstit{i}\phi$
                    \hfill for $k \geq 1$
\end{itemlist}
by the rules of modus ponens and $\cstit{i}$-necessitation.

\begin{pf}
If $\agtset $ is finite then
Sahlqvist's Theorem warrants that our axiomatics of Section \ref{sec:unsettling}
is sound and complete w.r.t.\ Kripke models satisfying the general permutation property.
We show in the annex that this can be extended to the infinite case.
\end{pf}
\end{theorem}

\goodbreak
\section{Complexity}\label{sec:complexity}

The axiom system of the preceding section allows us to characterize the complexity
of satisfiability of \STIT\ formulas.
We study separately the cases of Chellas' \STIT\ and of the 
deliberative \STIT.

\subsection{Complexity of Chellas' \STIT}

First, satisfiability of \CSTIT-formulas
can be decided in nondeterministic exponential time.

\begin{lemma}\label{lem:ComplexCstitUpper}
The problem of deciding satisfiability of a formula of \LCSTIT\ 
is in NEXPTIME.

\begin{pf}
This can be proved by the standard filtration construction,
which establishes that in order to know whether a formula $\phi$
is satisfiable in the Kripke models of Section \ref{sec:newSemantics}
it suffices to consider models having at most $2^{\lngth{\phi}} $ possible worlds.
See the annex for details.
\end{pf}

\end{lemma}

In the rest of the section we show that the upper bound is tight
if there are at least two agents. As usual we start with the two-agents case.

\begin{lemma}\label{lem:ComplexCstitLowerTwoAgents}
If $\card{\agtset} = 2$ then the problem of deciding satisfiability of
a formula of \LCSTIT\ is NEXPTIME-hard.

\begin{pf}
Remember our observation at the end of Section \ref{sec:unsettling}:
when $\card{\agtset} \ = 2$ then $\CSTIT_{\agtset} $
is nothing but the product logic S5$\otimes$S5.
We can then apply a result of Marx in \cite{Marx99}, who proved that the problem of
deciding membership of $\phi$ in S5$\otimes$S5 is NEXPTIME-hard.
(Actually Marx also proved membership in NEXPTIME.)
\end{pf}
\end{lemma}

Hence two-agent \CSTIT\ logic is NEXPTIME-complete.
Now we state NEXPTIME-completeness for any number of agents greater than $2$.

\begin{theorem}\label{theo:cstitNexptComplete}
If $\card{\agtset} \geq 2$ then the problem of deciding satisfiability
of a formula of \LCSTIT\ is NEXPTIME-complete.

\begin{pf}
See Annex.
\end{pf}

\end{theorem}

It remains to establish the complexity of single-agent \CSTIT.
It turns out that it has the same complexity as S5.

\begin{theorem}\label{theo:cstitSingleagentNpComplete}
If $\card{\agtset} = 1$ then the problem of deciding satisfiability
of a formula of \LCSTIT\ is NP-complete.

\begin{pf}
This can be proved by establishing an upper bound on the size of the models that is
quadratic in the length of the formula under concern.
\end{pf}

\end{theorem}

\begin{remark}
Intriguingly, while one-agent \STIT\ has the same complexity as S5, and
two-agent \STIT\ has the same complexity as S5$^2$,
$3$-agent \STIT\ \emph{does not} have the same complexity as S5$^3$:
while Xu's proof establishes decidability of \LCSTIT-formulas for any number of agents,
it was proved by Maddux that S5$^3$ is undecidable \cite{MarxMikulas00}.
\end{remark}

Thus we have characterized the complexity of satisfiability of \CSTIT\
formulas for all cases.

\goodbreak
\subsection{Complexity of the deliberative \STIT}

The complexity results for Chellas' \STIT\ do not immediately transfer to \DSTIT.
Indeed, the definition of the deliberative \STIT\ from the \CSTIT\ through
$[i\ \mathit{dstit}\! :\phi]$ = $\cstit{i} \phi] \land \lnot \Box \phi$
does not directly provide a lower bound for the deliberative \STIT\
because this is not a polynomial transformation.
We now establish these results by giving polynomial translations
from \CSTIT\ to \DSTIT and vice versa.

Let $\phi_0$ be any formula of \LDSTIT, and
let $\subfml(\phi_0) $ be the set of subformulas of $\phi_0$.
Let $\{p_\psi : \psi \in \subfml(\phi_0)\}$ be a set of (pairwise distinct) atoms none of
which occurs in $\phi_0$.
Every $p_\psi$ abbreviates the subformula $\psi $ of $\phi_0$.
We recursively define equivalences (`biimplications') that capture
the logical relation between $p_\psi$ and $\psi$.

\begin{definition}
We define:
\begin{center}
\begin{tabular}{lll}
$B_q$ &         = &  ($p_q \eqv q $)\\
$B_{\lnot \phi}$ & = & ($p_{\lnot \phi} \eqv \lnot p_{\phi} $)\\
$B_{\phi \land \psi}$& =  & ($p_{\phi \land \psi} \eqv p_{\phi} \land p_{\psi} $)\\
$B_{\Box \phi }$ & = & ($p_{\Box \phi} \eqv \Box p_{\phi} $)\\
$B_{[i:dstit \phi]}$ & = & ($p_{[i:dstit \phi]} \eqv
                                        [i]p_\phi \land \lnot \Box p_{\phi} )$
\end{tabular}
\end{center}
\end{definition}

\begin{definition}
We define the translation $tr $ from \DSTIT\ formulas to \CSTIT\ formulas as:
$tr(\phi_0) = p_{\phi_0} \land \bigwedge_{\psi \in \subfml(\phi_0)} \Box B_\psi $.
\end{definition}

\begin{theorem}\label{theo:dstitNexptHard}
$tr$ is a polynomial translation from \LDSTIT\ to \LCSTIT, and
for every formula $\phi_0$ of \LDSTIT,
$\phi_0$ is satisfiable iff $tr(\phi_0) $ is satisfiable.

\begin{pf}
See Annex.
\end{pf}
\end{theorem}

It follows that the problem of deciding whether a formula of \LDSTIT\ is
satisfiable is in NEXPTIME.
We now prove that this bound is tight.

\begin{definition}
We define equivalences $B'_\phi$ such that
$$B'_{\cstit{i}\phi}  =  (p_{\cstit{i}\phi} \eqv
                         [i\ \mathit{dstit}\! :p_\phi] \lor \Box p_{\phi} )$$
and $B'_\phi = B_\phi$ if $\phi$ is an atomic formula or if its main
logical connector is boolean.
\end{definition}

\begin{definition}
We define the translation $tr' $ from \LCSTIT\ to \LDSTIT\ as:
$tr'(\phi_0) =
p_{\phi_0} \land \bigwedge_{\psi \in \subfml(\phi_0)} \Box B'_\psi $.
\end{definition}

\begin{theorem}\label{theo:dstitInNexpt}
$tr'$ is a polynomial translation from \LCSTIT\ to \LDSTIT, and
for every formula $\phi_0$ of \LCSTIT,
$\phi_0$ is satisfiable iff $tr(\phi_0) $ is satisfiable.

\begin{pf}
The proof is analogous to that of Theorem \ref{theo:dstitNexptHard}.
\end{pf}
\end{theorem}

Together, Theorems \ref{theo:cstitNexptComplete}, \ref{theo:cstitSingleagentNpComplete},
\ref{theo:dstitNexptHard} and \ref{theo:dstitInNexpt} entail:

\begin{corollary}
The problem of deciding whether a formula of \LDSTIT\ is satisfiable is
NEXPTIME-complete if $\card{\agtset} \geq 2$, and
it is NP-complete if $\card{\agtset} = 1$.
\end{corollary}

\section{Conclusion}\label{sec:conclusion}

In this note we have established NEXPTIME-completeness of the satisfiability problem
of formulas of Chellas' \STIT\ and of the deliberative \STIT\
for the case of two or more agents.
All our complexity results appear to be new.

Our new axiom system for \STIT\ of Section \ref{sec:alterAx}
is an interesting alternative to Xu's.
It highlights the central role of the well-known equivalences
$\cstit{i} \cstit{j}  \phi  \eqv \Box \phi $ and
$\dstit{i} {\dstit{j} \phi} \eqv \bot $, for $i \not = j$
in theories of agency:
as we have shown, they allow to capture independence of agents just as
Xu's schema (AIA$_k$) does.

For the case of more than two agents, Section \ref{sec:unsettling}
provides a quite simple axiom system that is made up of very basic
modal principles, and moreover, does without historic necessity.

As we have pointed out in Section \ref{sec:alterAx},
an alternative axiomatics for the deliberative \STIT\ follows
straightforwardly.
We do not know whether the redundancy of historic necessity that we have established
for the \CSTIT\ in Section \ref{sec:unsettling} transfers to the deliberative \STIT.

\section*{Acknowledgements}

Thanks to Olivier Gasquet for comments and discussions.

\bibliographystyle{alpha}
\bibliography{biblio}

\addcontentsline{toc}{section}{Annex A: Proofs}
\section*{Annex: Proofs}

\subsection*{A.1: Proof of Lemma \ref{lem:validAaia}}

In order to prove the validity of every schema
\begin{itemlist}{(\InclBox{i})}
  \item[(AAIA$_k$)]
  $ \Diamond \phi \imp
    \poscstit{k} \bigwedge_{0 \leq i < k} \poscstit{i} \phi  $   \hfill for $k \geq 1$
\end{itemlist}
in BT+AC structures, we show that for every $w\in W$, $h,h' \in H_w$ and
$k \in \agtset$ there is $h_k \in Choice_k^w(h) $ such that
$h' \in Choice_i^w(h_k)$  for every $i \in \agtset \setminus \{k\}$.

Consider the strategy $s_w$ such that
$s_w(k) = Choice_k^w(h)$, and
$s_w(i) = Choice_i^w(h')$ for every $i \not = k$.
By the superadditivity constraint there is some $h_k$ such that
$h_k \in \bigcap_{i \in \agtset} s_w(i) $. Hence
$h_k \in Choice_k^w(h) $, and
$h'  \in Choice_i^w(h_k) $ for $i \not = k$.

\subsection*{A.2: Proof of Lemma \ref{lem:validGperm}}

We have to prove the validity of every schema
\begin{itemlist}{(\InclBox{i})}
\item[(GPerm$_k$)]
$\poscstit{l}\poscstit{m}\phi \imp
\poscstit{n}\bigwedge_{i\leq k, i \not = n} \poscstit{i}\phi$  \hfill for $k \geq 0$
\end{itemlist}
in BT+AC structures.

A look at the proof of Lemma \ref{lem:validAaia} shows that
$\Diamond \phi \imp
\poscstit{n}\bigwedge_{i\leq k, i \not = n} \poscstit{i}\phi$
is valid in BT+AC structures.
It therefore suffices to show the validity of
$\poscstit{l}\poscstit{m}\phi \imp \Diamond \phi $.
The latter is the case because
(1) $\poscstit{l}\poscstit{m}\phi \imp \Diamond \Diamond \phi $ is valid
(due to validity of axiom (\InclBox{i})), and
(2) $\Diamond \Diamond \phi \imp \Diamond \phi $ is valid
(due to validity of S5($\Box$)).

\subsection*{A.3: Proof of Theorem \ref{theo:Sahlqvist}}

We prove the theorem for the infinite case, i.e.\ $\card{\agtset} = \mathbb{N}$.
In this case the general permutation property is no longer a first-order property,
and Sahlqvist's result does not apply, i.e.\ the canonical model does not
necessarily satisfy the general permutation property.

Let $\phi$ be a formula that is consistent w.r.t.\ the axiomatic system
of Section \ref{sec:unsettling}.
Let $M = \tuple{W, R, V}$ be the canonical model associated to this system.
By arguments following the lines of those in the proof of Lemma \ref{lem:prop-rel} we have:
\begin{itemize}
\item $\forall i   \in \agtset$, $R_i$ is an equivalence relation;
\item $\forall i,j \in \agtset$ such that $i \not = j$, $R_i \circ R_j = R_1 \circ R_0$;
\item $(\bigcup_{i \in \agtset} R_i)^* = R_0 \circ R_1 = R_1 \circ R_0$.
\end{itemize}
By the truth lemma we may suppose that $M $ is generated via $R_1 \circ R_0$
from a possible world $w\in W$ such that $M,w \models \phi$.
Let $M' = \tuple{W', R', V'} $ be the filtration of $M$ w.r.t.\ $\subfml(\phi)$
(just as done in Annex A.4).
Note that $R_i' = W' \times W' $ for all $i\in \agtset $ not occurring in $\phi$.
This allows us to show that $M' $ satisfies the general permutation property.
From this completeness follows (via the filtration lemma).

\subsection*{A.4: Proof of Lemma \ref{lem:ComplexCstitUpper}}

Let  $M = \tuple{W,R,V}$ be a Kripke model such that
every $R_i$ is an equivalence relation and
$R$ satisfies the general permutation property.
Let $u$ be a world and $\phi$ a formula of \LCSTIT\ such that
$M,u\models\phi$.
Suppose that $M$ is generated from $w$ through $R_1 \circ R_0$.
(This can be supposed w.l.o.g.\ because of Lemma \ref{lem:prop-rel}
of Section \ref{sec:newSemantics}.)
$\subfml(\phi)$ being the set of all subformulas of $\phi$,
we say $w$ and $v$ are $\subfml(\phi)$-equivalent iff
$\forall \psi \in \subfml(\phi),\ (M,w\models\psi \mbox{ iff }
M,v \models\psi )$, and note $w \equiv_{\subfml(\phi)} v$.
Let $\ext{w}_{\equiv_{\subfml(\phi)}} $ denote
the equivalence class of $w$ modulo $\equiv_{\subfml(\phi)}$.

We construct $M' = \tuple{W',R',V'}$ such that:
\begin{itemize}
\item $W' = W|_{\equiv_{\subfml(\phi)}}
          = \{ \ext{w}_{\equiv_{\subfml(\phi)}} : w \in W \}$
\item $\tuple{\ext{w},\ext{v}} \in R'_i $        iff
      $\forall \cstit{i}\psi \in \subfml(\phi),\
       (M,w \models\cstit{i}\psi \mbox{ iff } M,v\models\cstit{i}\psi)$
\item $V'(p) = \{\ext{w}\ : w \in V(p) \}$ for all $p \in \subfml(\phi)$
\end{itemize}
Remark that for all $i \in \agtset$,
if $i$ does not occur in $\phi$ then $R_i' = W' \times W' $.

We must check that every $R'_i$ is an equivalence relation,
that $M'$ verifies the general permutation property,
that for all $\psi \in \subfml(\phi)$ and $w \in W$,
$M,w \models \psi$ iff $M',\ext{w} \models \psi$, and
that $\card{W'}$ is exponential in the length of $\phi$:
\begin{enumerate}
\item Every $R'_i$ is an equivalence relation, and
$M'$ satisfies the general permutation property.

This follows from the definition of $R'_i$.

\item $\forall \psi \in\subfml(\phi), \forall w\in W,\ (M,w\models \psi
\mbox{ iff } M',\ext{w}\models \psi)$.

This follows from the filtration lemma (see \cite{Blackburn:2001:ML} for details).

\item $\card{W'} \leq 2^{\lngth{\phi}}$

Note that members of $W'$ are subsets of states of $W$ satisfying
exactly the same formulas of $\subfml(\phi)$.
Thus $\card{W'} \leq 2^{\card{\subfml(\phi)}}$ corresponding to the set
of subsets of $\subfml(\phi)$.
We can show by induction on $\psi$ that $\card{\subfml(\psi)} \leq \lngth{\psi}$
and then conclude.

\end{enumerate}

Hence, $\forall \phi \in $ \LCSTIT,
if $\phi$ is satisfiable then
$\exists M = \tuple{W,R,V}$ such that $\card{W} \leq 2^{\lngth{\phi}}$ and
there is $w\in W$ such that $M,w\models\phi$.
It allows us to propose a decision procedure with input $\phi \in $ \LCSTIT,
and which works as follows:
guess an integer $N \leq 2^{\lngth{\phi}}$ and a model $M = \tuple{W,R,V}$
such that $\card{W} \leq N$;
then check whether there is a $w \in W$ such that $M,w \models \phi$.

\subsection*{A.5: Proof of Theorem \ref{theo:cstitNexptComplete}}

The upper bound is given by Lemma \ref{lem:ComplexCstitUpper}.

To establish the lower bound consider 
the set of formulas where only the agent symbols $0$ and $1$ occur.
We show that deciding satisfiability of any formula of that fragment
is NEXPTIME-hard, for any $\agtset$ such that $\card{\agtset} \geq 2$.
If $\agtset $ is just $\{0,1\} $ this holds
by Lemma \ref{lem:ComplexCstitLowerTwoAgents}.
Else we prove that if $\{0,1\} \subset \agtset $ then
the logic of Kripke models for $\agtset$ is a conservative extension of
that for $\{0,1\}$.

Let $\phi$ be any formula containing only $0$ and $1$.

For the left-to-right direction, suppose $\phi$ is valid in all Kripke models
for the set of agents $\{0,1\} $.
By Theorem \ref{cool-theo}, $\phi $ can then be proved from
axioms (GPerm$_1$), (Perm01), S5($0$) and S5($1$) with
the rules of modus ponens, $\cstit{0}$- and  $\cstit{1}$-necessitation.
Therefore $\phi$ is also provable from the `bigger' axiomatics for $\agtset$.

For the right-to-left direction, suppose there is a Kripke model
$M = \tuple{W, R, V} $ for the set of agents $\{0,1\}$ and a $w \in W $ such that
$M,w \models \phi $, where $R : \{0,1\} \longrightarrow \mathcal{P}(W \times W)$
associates to every $i \in \{0,1\}$ an equivalence relation $R_i$ on $W$.
We are going to build a Kripke model $M'$ for the bigger set of agents $\agtset$
such that $M',w \models \phi$.
Let $M' = \tuple{W, R', V} $ such that
$R' : \agtset \longrightarrow \mathcal{P}(W \times W)$ with
$R'_0 = R_0 $,
$R'_1 = R_1 $ and
$R'_i = R_0 \circ R_1 $ for $i \geq 2$.
Clearly $M',w \models \phi$, too.
It remains to show that $M' $ is indeed a Kripke model
as required in Section \ref{sec:newSemantics}.
By item $2$ of Lemma \ref{lem:prop-rel}
every $R'_i $ is an equivalence relation, so we only have to show that
the general permutation property holds in $M'$:
if $\tuple{w,v} \in R'_l \circ R'_m $ then
there is $u_n \in W$ such that:
$\tuple{w,u_n} \in R'_n $ and $\tuple{u_n,v} \in R'_i $
for every $i \in \agtset \setminus \{n\} $
(cf.\ Lemma \ref{lem:validGperm}).
First we show that for every $l$ and $m$
we have $R'_l \circ R'_m = R_0 \circ R_1$.
\begin{itemize}
\item If $i=0$ and $j=1$ then trivially
$R'_l \circ R'_m = R_0 \circ R_1$.
\item If $l=1$ and $m=0$ then
$R'_l \circ R'_m = R_1 \circ R_0
                 = R_0 \circ R_1 $
\item If $l=0$ and $m\geq 2$ then
$R'_l \circ R'_m = R_0 \circ R_0 \circ R_1
                 = R_0 \circ R_1 $
\item If $l=1$ and $m\geq 2$ then
$R'_l \circ R'_m = R_1 \circ R_0 \circ R_1
                 = R_0 \circ R_1 \circ R_1
                 = R_0 \circ R_1 $
\item If $l\geq 2$ and $m=0$ then
$R'_l \circ R'_m = R_0 \circ R_1 \circ R_0
                 = R_0 \circ R_0 \circ R_1
                 = R_0 \circ R_1  $
\item If $l\geq 2$ and $m=1$ then
$R'_l \circ R'_m = R_0 \circ R_1 \circ R_1
                 = R_0 \circ R_1  $
\item if $l\geq 2$ and $m\geq 2$ then
$R'_l \circ R'_m = R_0 \circ R_1 \circ R_0 \circ R_1
                 = R_0 \circ R_0 \circ R_1 \circ R_1
                 = R_0 \circ R_1  $
\end{itemize}
(The identities in all these items hold because
$R_0$ and $R_1$ permute by item $1$ of Lemma \ref{lem:prop-rel}, and
because $R_0 $ and $R_1 $ are equivalence relations.)
Thus $\tuple{w,v} \in R'_l \circ R'_m $ implies $\tuple{w,v} \in R_0 \circ R_1 $.
We have to show that for every $n \geq 1$ there is $u_n \in W$ such that:
$\tuple{w,u_n} \in R'_n $ and $\tuple{u_n,v} \in R'_i$, for every $i \in \agtset $.
\begin{itemize}
\item
For $n=1$, $\tuple{w,v} \in R_0 \circ R_1 $ implies that
$\tuple{w,v} \in R_1 \circ R_0$ by item $1$ of Lemma \ref{lem:prop-rel},
and the latter implies that $\tuple{w,v} \in R'_1 \circ R'_0 $.
Therefore there is a $u_1$ such that $\tuple{w,u_1}\in R'_1 $ and $\tuple{u_1,v} \in R'_0 $.
\item
For $n\geq 2$, take $u_n=v$:
$\tuple{w,v} \in R_0 \circ R_1 $ implies that $\tuple{w,v} \in R'_n $ by definition of $R'_n$, and
we have $\tuple{v,v} \in R'_i $ because every $R'_i $ is an equivalence relation
(for $i \geq 2$ this is the case by item $2$ of Lemma \ref{lem:prop-rel}).
\end{itemize}

\subsection*{A.5: Proof of Theorem \ref{theo:dstitNexptHard}}

The proof is done via the following lemmata.

\begin{lemma}\label{lem:dsat2trcsat}
For all formulas $\phi_0 $ in the language of \DSTIT,
if   $   \phi_0  $ is satisfiable
then $tr(\phi_0) $ is satisfiable.

\begin{pf}
Suppose there is $M = \tuple{W,R_\Box,R,V} $ such that  $M,w \models \phi_0 $.
We build a model $M' = \tuple{W,R_\Box,R,V'} $ such that $M',w \models tr(\phi_0) $
by setting
$V'(q) = V(q) $ for all atoms $q $ appearing in $\phi_0$, and
$V'(p_\psi) = \{w\in W : M,w\models \psi \} $
for all $\psi \in \subfml(\phi_0) $.

By induction on the structure of $\psi$ we show that $M,v \models B_\psi $
for all $v \in W$ and all $\psi \in \subfml(\phi_0) $.
(Details left to the reader.)

Hence $M' \models \bigwedge_{\psi \in \subfml(\phi_0)}      B_\psi $, and also
      $M' \models \bigwedge_{\psi \in \subfml(\phi_0)} \Box B_\psi $.
Since $M,w \models \phi_0$, we have $M',w \models p_{\phi_0} $ by construction of $V'$.
Thus  $M',w \models p_{\phi_0} \land \bigwedge_{\psi \in \subfml(\phi_0)} \Box B_\psi $,
in other words
      $M',w \models tr(\phi_0) $.
\end{pf}
\end{lemma}

\begin{lemma}\label{lem:trdsat2csat}
For all formulas $\phi_0 $ in the language of \DSTIT,
if   $tr(\phi_0) $ is satisfiable
then $   \phi_0  $ is satisfiable.

\begin{pf}
Suppose there is $M = \tuple{W,R_\Box,R,V} $
such that  $M,w \models tr(\phi_0) $. Thus
$M,w \models p_{\phi_0} \land \bigwedge_{\psi \in \subfml(\phi_0)} \Box B_\psi $.
By induction on the structure of $\psi$ we show that
$M,v \models p_\psi \eqv \psi $
for all $v \in W$ and all $\psi \in \subfml(\phi_0) $.
(Details left to the reader.)

Thus  $M,w \models p_{\phi_0} $, and $M,w \models p_{\phi_0} \eqv \phi_0 $.
Hence $M,w \models    \phi_0  $.
\end{pf}
\end{lemma}

\begin{lemma}
$tr$ is a polynomial transformation.

\begin{pf}
We easily show that $\lngth{B_\psi} \leq 12$ and
$\lngth{ \bigwedge_{\psi \in \subfml(\phi_0)} \Box B_\psi }
\leq
\lngth{\phi_0} . (2+\lngth{B_\psi})$.
Then, $\lngth{\bigwedge_{\psi \in \subfml(\phi_0)}\Box B_\psi} \leq
14.\lngth{\phi_0} $.
We conclude that $\lngth{tr(\phi_0)}  \leq 1+14.\lngth{\phi_0}$.
Remark that $\card{\subfml(\phi_0)} \leq \lngth{\phi_0} $.
Moreover, for every formula $\phi$ in the language of \CSTIT,
$\lngth{B_\phi} = {\cal O}(\lngth{\phi}) $.
As a result,
$\lngth{tr(\phi_0)} = {\cal O}(\lngth{\phi_0}^2) $.

\end{pf}
\end{lemma}

\end{document}